\documentclass[aps,prl,preprint ,groupedaddress]{revtex4-1}

\usepackage{graphicx}
\usepackage{amsmath,amsfonts,amssymb}
\usepackage[colorlinks=true,linkcolor=blue,citecolor=blue]{hyperref}
\usepackage{CJK}

\begin{document}

\begin{CJK*}{Bg5}{bsmi}
\title{Gate tuneable beamsplitter in ballistic graphene}


\author{Peter Rickhaus}
\affiliation{Department of Physics, University of Basel, Klingelbergstrasse 82, CH-4056 Basel, Switzerland}

\author{P\'eter Makk}
\affiliation{Department of Physics, University of Basel, Klingelbergstrasse 82, CH-4056 Basel, Switzerland}
\email{Peter.Makk@unibas.ch}

\author{Ming-Hao Liu}
\affiliation{Institut f\"ur Theoretische Physik, Universit\"at Regensburg, D-93040 Regensburg, Germany}

\author{Klaus Richter}
\affiliation{Institut f\"ur Theoretische Physik, Universit\"at Regensburg, D-93040 Regensburg, Germany}

\author{Christian Sch\"onenberger}
\affiliation{Department of Physics, University of Basel, Klingelbergstrasse 82, CH-4056 Basel, Switzerland}

\date{\today}

\begin{abstract}
We present a beam splitter in a suspended, ballistic, multiterminal, bilayer graphene device. By using local bottomgates, a p-n interface tilted with respect to the current direction can be formed. We show that the p-n interface acts as a semi-transparent mirror in the bipolar regime and that the reflectance and transmittance of the p-n interface can be tuned by the gate voltages. Moreover, by studying the conductance features appearing in magnetic field, we demonstrate that the position of the p-n interface can be moved by $1\,\mu$m. The herein presented beamsplitter device can form the basis of electron-optic interferometers in graphene.
\end{abstract}


\maketitle
\end{CJK*}


Semi-transparent mirrors act as beam splitters in optical experiments. They are important building blocks for many interference experiments, be it a Fabry-P\'{e}rot, a Michelson or a Mach-Zehnder two path interferometer. In two-dimensional electron gases (2DEGs) such mirrors have been constructed using quantum point contacts in the quantum Hall regime. Thereby, the Mach-Zehnder experiment could be implemented \cite{Ji2003} involving, however, strong magnetic fields.

Graphene offers the unique possibility to mimic optical systems once transport is ballistic. Due to recent advances in fabrication techniques, ballistic electron transport can be observed on the micrometer scale as demonstrated by magnetic focusing experiments \cite{Mayorov2011, Taychatanapat2013, Calado_APL}.
By using p-n interfaces, Fabry-P\'erot interferometers have been realized in single-layer \cite{Young2009,Rickhaus2013,Grushina2013}, gapped bilayer \cite{Varlet2014} and trilayer graphene \cite{Campos2012}. Moreover, the observation of electron guiding, snake states \cite{Rickhaus2015, Taychatanapat2015} or ballistic supercurrents \cite{Calado2015, Allen2015, Shalom2015} highlighted the possibilities of p-n junctions in graphene.

P-n interfaces formed in graphene can be reflective, transparent or semi-transparent, depending on the angle of incidence of the charge carriers and the shape of the potential that forms the interface. For smooth p-n junctions, trajectories close to zero incidence angle are transmitted as a result of Klein tunneling, whereas electrons arriving under large angles are reflected. This suggests that by using a tilted p-n interface, where the Klein-tunneling trajectories are not dominating, a partially transparent mirror can be achieved. In fact, measurements on short and tilted p-n interfaces in graphene devices on SiO$_2$ revealed an increase in two-terminal resistance \cite{Sutar2012,Sajjad2012}.

Here we present the realization of a semi-transparent mirror in suspended graphene, using a bottomgate structure which is tilted with respect to the current flow direction. The presented four-terminal device allows us to measure reflectance and transmission of the mirror in a ballistic, ungapped bilayer sample. We show that in the unipolar regime, the measured currents can be understood within a simple geometrical picture, whereas in the bipolar regime a partially reflective mirror is formed. Moreover, we demonstrate that the transport properties in weak magnetic field can be substantially altered by moving the position of the mirror by distances up to $1\,\mu$m. Finally we discuss possibilities for the realization of future graphene interferometers based on the present device.


\begin{figure}[htbp]
    \centering
      \includegraphics[width=.6\columnwidth]{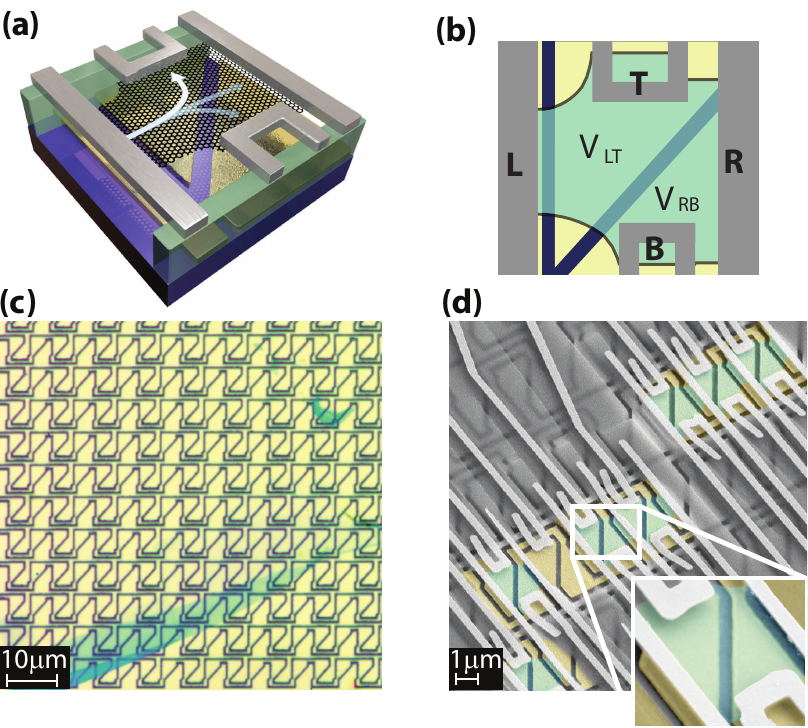}
    \caption{
	\textbf{a-b}, Three-dimensional design and schematic representation of the mirror device. The Pd-contacts are gray, the bottomgates golden and the LOR pillars are colored green. The tilted bottomgate structure allows to form an oblique p-n interface. Electrons injected at the L contact will be either reflected towards T or transmitted towards the R or B contact.
	\textbf{c}, Optical image of an area covered with bottomgate structures. The turquoise parts are few-layer graphene on top of LOR.
	\textbf{d}, Scanning electron micrograph with a zoom-in window. Graphene is colored turquoise here, and the LOR resist is gray and semi-transparent.
\label{fig:mirror1}}
\end{figure}


In order to measure the reflectance of a bilayer p-n interface we designed a four-terminal sample as shown in Figure \ref{fig:mirror1}a. Bilayer graphene is expected to exhibit a more reflective p-n interface \cite{Katsnelson2006,Liu22012}, due to anti-Klein tunnelling. The contacts and gates are labeled in the schematic top-view of Figure \ref{fig:mirror1}b. Using the two bottomgates $V_{\rm LT}$ and $V_{\rm RB}$, a tilted p-n interface can be formed and the reflectance of the mirror can be studied.

We fabricate the samples using a resist-based suspension technique as described  in detail in Refs.~\cite{Tombros2011, Maurand2014}. First, a large area of tilted bottomgate structure is prefabricated on undoped Si substrate and spin-coated with lift-off resist (LOR). Afterwards, graphene is transferred on top of the gate array. Due to the large patterned area, no special care needs to be taken during alignment. In Figure \ref{fig:mirror1}c an optical image of such a bottomgate array after LOR and graphene deposition is shown. The bottomgate array is tuned by three voltages, allowing to influence two interfaces for each device independently - a non-tilted interface close to the L contact and the tilted mirror interface. However, since the first interface is very close to the L contact for the measured sample, we did not see a change in the transport characteristics using this gate. For simplicity we therefore connect this gate to the first tilted gate and refer to it as $V_{\rm LT}$ in the following.
The turquoise parts in Figure \ref{fig:mirror1}c are few-layer graphene flakes ($>3$ layers). Thinner flakes are not visible in the optical microscope after transfer, but their positions are known from images recorded before.

In a further step, graphene is etched in oxygen plasma, contacted with palladium (Pd) contacts and suspended using e-beam exposure and subsequent development of LOR. An SEM image after suspension is depicted in Figure \ref{fig:mirror1}d. Afterwards, the graphene is cleaned by current-annealing at low temperatures. We note that the U-shaped side contacts are mechanically stable during current annealing and that the large distance between bottomgate and graphene allows to tune the position of the mirror by $1\,\mu$m as we will reveal later.
The measurements are done by standard Lock-In technique, where a small AC voltage is applied e.~g. at the left (L) contact and current is recorded at the other terminals separately.


\begin{figure}[h]
    \centering
      \includegraphics[width=.56\columnwidth]{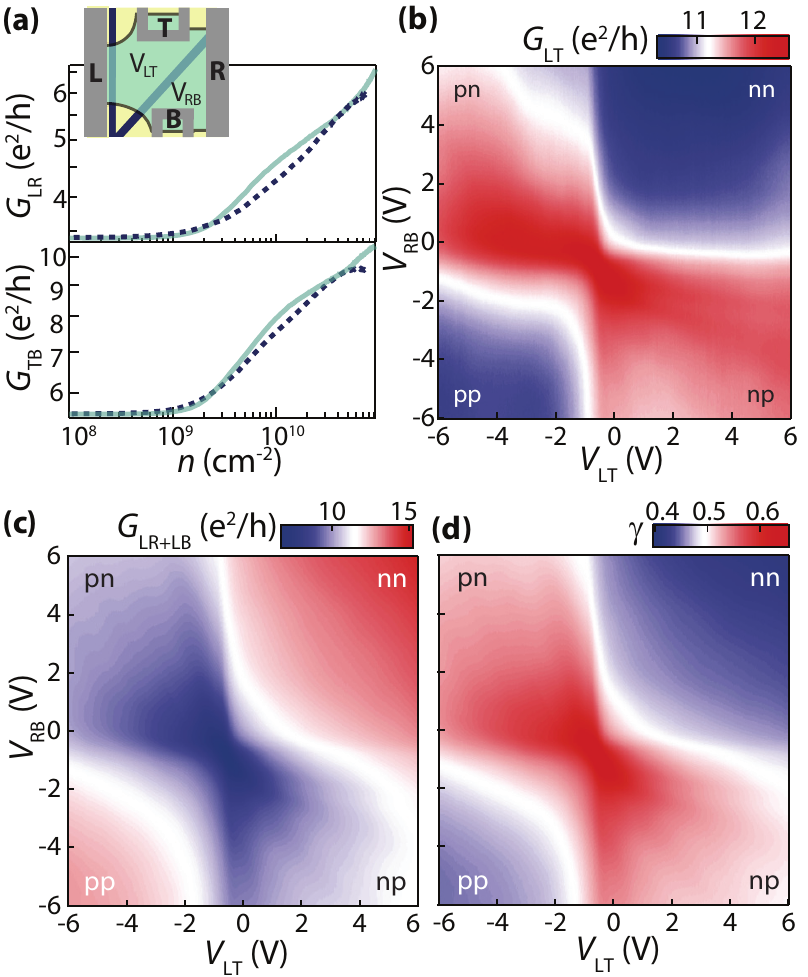}
    \caption{
    \textbf{a}, Conductances $G_{\rm LR}$ and $G_{\rm BT}$ across the device as a function of unipolar gating ($V_{\rm LT}=V_{\rm RB}$) reveal residual doping of $n_0\approx 1...2\cdot10^9\,$cm$^{-2}$ for the electron (turquoise) and hole-side (dashed).
	\textbf{b}, $G_{\rm LT}$ measured between L and T contact, as a function of the mirror gates $V_{\rm LT}$ and $V_{\rm RB}$ showing an increased conductance in the bipolar regime.
	\textbf{c}, In contrast, the conductance from the L to the R and B contact is higher in the unipolar regime.
	\textbf{d}, The reflectance of the mirror $\gamma=I_{\rm LT}/I_{\rm tot}$ is increased from 40\% in the unipolar to above 60\% in the bipolar regime.
\label{fig:mirror2}}
\end{figure}

For characterization after current annealing we measure the conductance across the device, i.~e. from L to R ($G_{\rm LR}$) and from T to B ($G_{\rm TB}$), as a function of unipolar gate tuning ($V_{\rm LT}=V_{\rm RB}$). These field effect measurements are shown in Figure \ref{fig:mirror2}a to reveal the residual doping $n_0$ after current annealing. The traces flatten in the range of $n_0\approx 1...2\cdot10^9\,$cm$^{-2}$ for  the electron (turquoise) and hole (blue dashed) doping in both directions across the device, proving the high quality of the measured device. We further extracted the field effect mobility $\mu=d\sigma/dn\times 1/e\approx 120'000\,\textrm{cm}^2\textrm{V}^{-1}\textrm{s}^{-1}$. Considering comparable devices where ballistic transport has been explicitly demonstrated \cite{Rickhaus2013,Grushina2013} these numbers suggest that the transport is dominated by ballistic trajectories.

In Figure \ref{fig:mirror2}b the conductance $G_{\rm LT}(V_{\rm LT},V_{\rm RB})$ is shown. Upon the formation of a p-n interface, more charge carriers will reach the top contact and the conductance is increased in the p-n and n-p regions (red) compared to the unipolar p-p or n-n situation (blue). For the transmitted charge carriers reaching the R and B contact ($G_{\rm LR+LB}(V_{\rm LT},V_{\rm RB})$) the conductance is lowered when the p-n interface is present (Figure \ref{fig:mirror2}c). Finally, the reflectance of the mirror is given by $\gamma=I_{\rm LT}/I_{\rm tot}$ with $I_{\rm tot}=I_{\rm LT}+I_{\rm LR}+I_{\rm LB}$ which is plotted in Figure \ref{fig:mirror2}d.
For uniform (n-n or p-p) gating roughly $\gamma_0:=\gamma(6\,{V},6\,{V})=40\%$ of the current reaches the T contact. Upon the formation of a p-n interface, $\gamma$ increases to 60\%. Individual maps $G_{\rm LR}$ and $G_{\rm LB}$ are given in the supporting information \cite{supporting}. The tuning of  $\gamma$ with local gate voltages shows that the device can be operated as gate-tuneable beam-splitter, and 50\%-50\% splitting, can be achieved. In two-path interferometers, the 50\%-50\% splitting is used usually, since this maximizes the visibility of the interference signals.

\begin{figure}[htbp]
    \centering
      \includegraphics[width=.6\columnwidth]{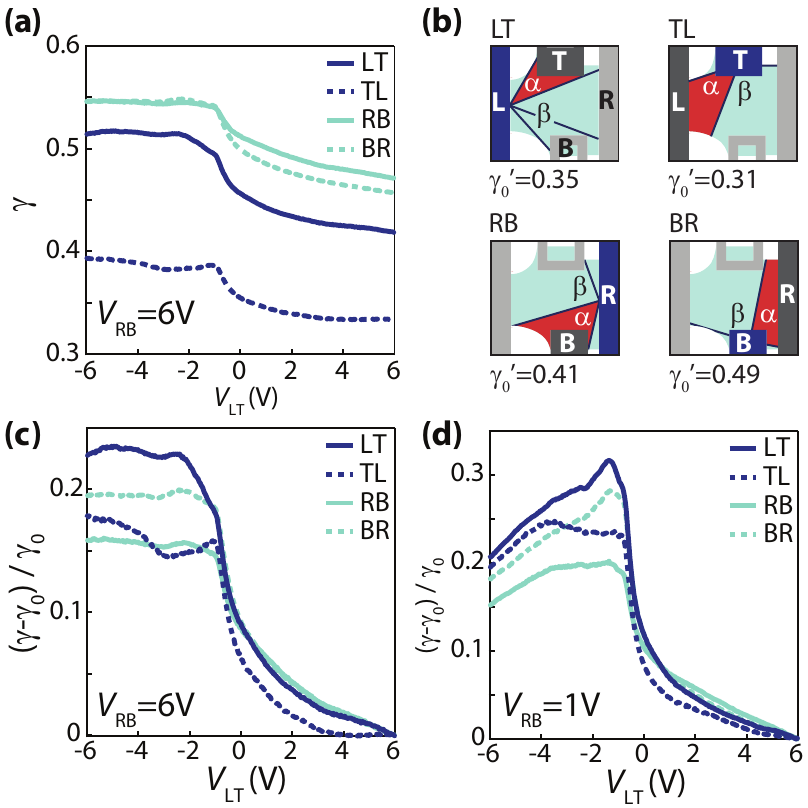}
    \caption{
    \textbf{a}, $\gamma(V_{\rm LT})$ at $V_{\rm RB}=6\,V$ for different device configurations. The dark blue curve for instance corresponds to a cut in the $\gamma_{\rm LT}$ colorscale plot of Figure \ref{fig:mirror2}d.
     \textbf{b}, Different measurement configurations. The injector contact is colored blue.
    \textbf{c}, Absolute increase of reflectance $(\gamma-\gamma_0)/\gamma_0$, where $\gamma_0=\gamma((6,6)\,V)$ is a geometrical factor.
    \textbf{d}, Similar plot for $V_{\rm RB}=1\,V$.\label{fig:mirror3}
    }
\end{figure}

We further investigate the reflection properties of the p-n interface by recording the reflected conductances for different injector contacts. The reflectance in different measurement configurations (explained in Fig.~\ref{fig:mirror3}b) is shown in Figure~\ref{fig:mirror3}a, where curves of $\gamma(V_{\rm LT},V_{\rm RB}=6\,{V})$ are plotted. The blue curve corresponds to a cut in the colorscale plot of Figure \ref{fig:mirror2}d. For the blue dashed curve, current is injected at the T contact and $\gamma$ is given by $I_{\rm TL}/(I_{\rm TL}+I_{\rm TB}+I_{\rm TR})$. In a corresponding way the (dashed) turquoise line corresponds to injection at the R (B) contact. As before, $\gamma_0$ is the reference reflectance, without interface. The obtained reference reflectances in the unipolar regime are roughly consistent with a simple geometric consideration, sketched in the schematics of Figure \ref{fig:mirror3}b. Ballistic charge carriers, injected from the middle of the L contact, reach the T contact under a solid angle of $\alpha$ and the R and B contacts under $\beta$. The ratio $\gamma_{0\rm,LT}'=\alpha/(\alpha+\beta)=0.35$ is roughly consistent with the measured $\gamma_{0\rm,LT}=0.41$. This holds also for the ratio of current reaching the R or B contact, i.e. $\gamma_{0\rm,LR}'=0.34$ and $\gamma_{0\rm,LR}=0.26$ and similarly $\gamma_{0\rm,LB}'=0.31$ and $\gamma_{0\rm,LB}=0.33$. Furthermore, the ratios for different measuring configurations are $\gamma_{0\rm,TL}'=0.31$ and  $\gamma_{0\rm,TL}=0.33$, $\gamma_{0\rm,RB}'=0.41$ and  $\gamma_{0\rm,RB}=0.48$,$\gamma_{0\rm,BR}'=0.41$ and  $\gamma_{0\rm,BR}=0.47$. Deviations are due to the strong simplification of the model, contact doping and varying contact resistance.

In Figure~\ref{fig:mirror3}c, the relative increase of reflectance $(\gamma-\gamma_0)/\gamma_0$ for $V_{\textrm{RB}}=6\,$V is shown. The highest value is reached for injection at the L contact. The device is designed for this configuration since direct trajectories from L to T or B are minimized.
The highest values of reflectance can be found close to the CNP as can be seen in Fig.~\ref{fig:mirror2}b and in Fig.~\ref{fig:mirror3}d, which shows the relative increase of the reflectance for $V_{\textrm{RB}}=1\,$V. We think that this is the result of short-cut currents flowing at the edges prominent at low densities. First, the electric field at the sample edge is larger, leading to increased doping at the edges, since the bottomgate structure extends much further than the flake. Second, residual dopants tend to accumulate close to the contacts after current annealing \cite{Freitag2012b}, also leading to currents that remain unaffected by the formation of a p-n interface. And third, the doping of the contacts becomes more significant. These effects lead to larger relative currents at the edges compared to the bulk, and these currents have more relative weight at low densities. The effect of these edge currents are prominent for currents flowing from L to T and L to B, whereas it is reduced for currents from L to R (as seen in the maps of the supporting info \cite{supporting}), which results in the increase seen in Figure~\ref{fig:mirror3}d.

Even if these currents could be drastically reduced, an efficiency of $100\%$ cannot be achieved in our device, since electrons reach the (bilayer) p-n interface under a wide range of angles. This is the result of extended contact size and also of the lack of collimation. Some of these trajectories reaching the interface will always have a finite transmission \cite{Katsnelson2006}. These trajectories have small, but non-zero incidence: at zero incidence the transmission is zero (anti-Klein tunneling) and by increasing the angle a finite transmission probability becomes possible, but the smoothness of the junction leads to an exponential suppression for larger angles \cite{Cheianov2006}.

\begin{figure}[htbp!]
    \centering
      \includegraphics[width=.55\columnwidth]{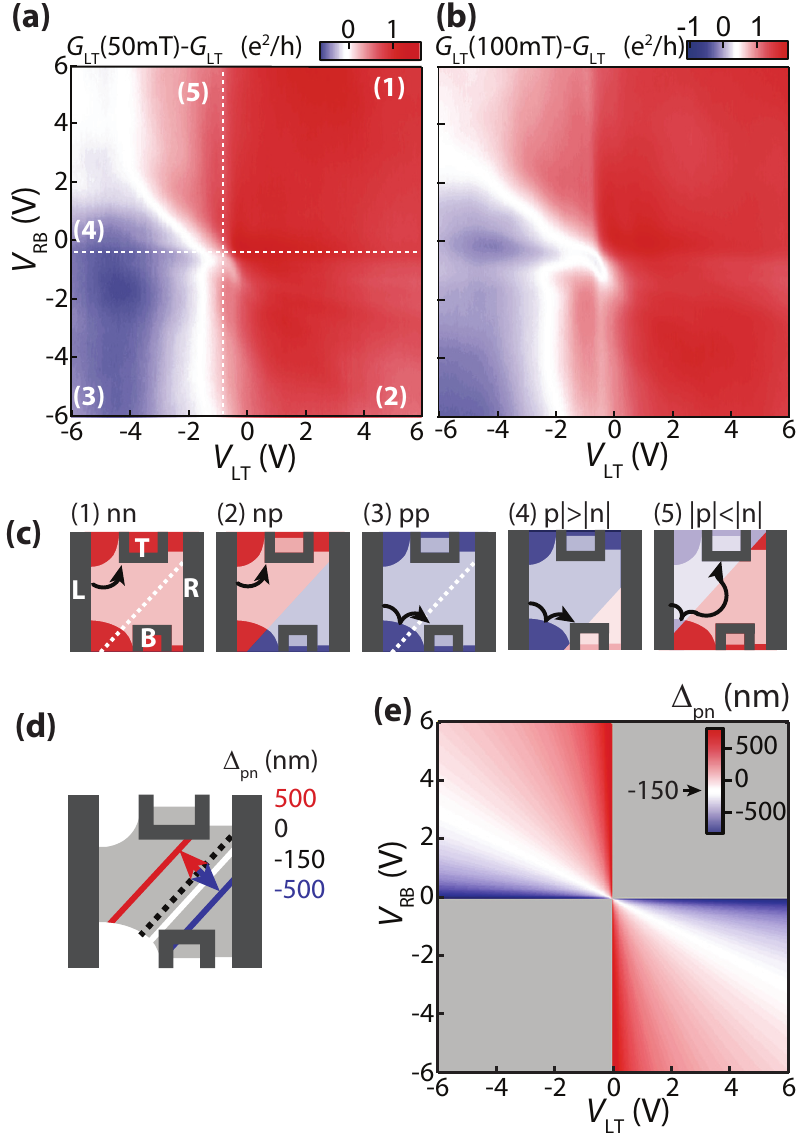}
    \caption{
	\textbf{a}, Increase of the mirrored signal in magnetic field of 50$\,$mT: $G_{\rm LT}(B=50\,$mT$)- G_{\rm LT}(B=0\,$T). The plot shows a strong asymmetry as a function of $V_{LT}$, which is due to bent trajectories in B field. The upper left quadrant is split into an enhanced and a reduced part along a diagonal line.
	\textbf{b}, Similar plot showing the difference of $G_{\rm LT}$ at $100\,$mT an $0\,$T.
	\textbf{c}, The sketches explain the conductance regions seen in panel a) and b).
	\textbf{d}, Geometry of the device showing the relative position of the p-n interface for different $\Delta_{\rm pn}$ values.
	\textbf{e}, Colorscale map from electrostatic simulations, revealing that $\Delta_{\rm pn}$ depends on the ratio of $V_{\rm LT}/V_{\rm RB}$. Our device offers a high tuning range ($\pm 500\,$nm) due to the large distance between bottomgate and graphene. White color corresponds to $-150\,$nm.
\label{fig:mirror4}}
\end{figure}

By applying a perpendicular magnetic field $B$, the amount of electrons reaching the T contact (injected from L) can be increased using magnetic focussing. This is seen in Figure \ref{fig:mirror4}a and b, where we show the conductance increase with respect to zero field measurement, i.~e. $G_{\rm LT}(50\,$mT$)-G_{\rm LT}(0\,$T$)$ and $G_{\rm LT}(100\,$mT$)-G_{\rm LT}(0\,$T$)$, respectively.
The structure of the maps is explained using the sketches in Figure \ref{fig:mirror4}c. In the case of unipolar n-n doping, in region (1) in Fig.~\ref{fig:mirror4}a, $G_{\rm LT}$ rises (by $\approx1\,{e^2/h}$) since the electrons are deflected towards the T contact by the Lorentz force, as shown in the corresponding sketch of Fig.~\ref{fig:mirror4}c. The cyclotron diameter at $50\,$mT is with $1.4\,\mu$m at $V_{\rm LT}=V_{\rm RB}=6\,$V in the range of the geometrical dimensions (the distance between L and T is $1.1\,\mu$m), implying  that we are in situation of magnetic focusing \cite{Houten1988,Taychatanapat2013}. A clear focusing signal is however not expected due to the large size of the contacts. The increase is most pronounced at small gate voltages, where the cyclotron radius is smallest. The additional current at the T contact is not influenced if an n-p interface is formed by lowering $V_{\rm RB}$, as sketched in Figure \ref{fig:mirror4}c (2). For this reason, the conductance increases similarly in regions (1) and (2), as seen in Fig.~\ref{fig:mirror4}a. However, if $V_{\rm LT}$ is decreased, the cyclotron motion changes sign once the polarity of charge carriers is inverted (3). In this p-p region, the holes are deflected towards the B contact, leading to a decreased conductance measured at T. This decrease persists in the p-n region (4) for large $|V_{\rm LT}|$ and low $|V_{\rm RB}|$. But surprisingly, the conductance is enhanced in the opposite case (5), i.e. low $|V_{\rm LT}|$ and large $|V_{\rm RB}|$.

The structure in the p-n region in  Fig.~\ref{fig:mirror4}a can be understood by considering two effects. First, once a p-n interface is present, current flows along this interface and the formation of snake states is expected. Recently snake states were observed in single layer graphene  \cite{Rickhaus2015,Taychatanapat2015}, and for bilayer graphene also an increased current along the interface is expected \cite{bilayersnake}. The second effect takes into account the large distance between graphene and the bottomgates ($600\,$nm) that allows to change the position $\Delta_{\rm pn}$ of the p-n interface drastically by the gate voltage. In Figure \ref{fig:mirror4}d the position of the p-n interface for symmetric gating is drawn as a black dashed line. By lowering the density in the LT cavity (i.e. by lowering $|V_{\rm LT}|$) the interface can be shifted up to $500\,$nm towards the T contact (red line). On the other hand, by decreasing $|V_{\rm RB}|$, the interface is shifted towards the B contact by a similar amount (blue line). A colorscale map revealing $\Delta_{\rm pn}(V_{\rm LT},V_{\rm RB})$ is given in Figure \ref{fig:mirror4}e. The map was calculated using the method described in Ref.~\cite{Liu2015} with the geometry of our device, neglecting quantum capacitance. The white region marks the transition between region (4) and (5) in Figures \ref{fig:mirror4}b,c and corresponds to $\Delta_{\rm pn}=-150\,$nm. The corresponding position of the p-n interface is sketched as a white line in Figure \ref{fig:mirror4}d, where the interface crosses the B contact. For larger $|V_{\rm RB}|/|V_{\rm LT}|$, the presence of the p-n interface is negligible for the injected current in L, as sketched in Figure \ref{fig:mirror4}c (4), explaining the similarity to region (3). However, the interface transports charge carriers in direction of the T contact in the opposite case (5), leading to an increased $G_{\rm LT}$ at $50\,$mT and $100\,$mT.




The device discussed here presents the realization of a semi-transparent graphene mirror with movable position. This device can be the fundamental building block of a Michelson-Morley or a Mach-Zehnder interferometer. As an important improvement for the realization of such interfereometers, a collimator interface can be added. The strong collimation offered by smooth p-n interfaces \cite{Cheianov2006} can be harvested to create a plane wave in graphene. The reflective mirrors of the optical system can be replaced either by edges of the graphene flake, (reflective) contacts or additional p-n interfaces. The reflection at the graphene edges is mostly specular rather than diffusive, as has been demonstrated by magnetic focusing experiments \cite{Taychatanapat2013, Calado_APL}. The above demonstrated beam-splitter lies at the heart of these two path interferometers and brings cross-correlation measurements on ballistic graphene interferometers within reach.

\textbf{Acknowledgments}

This work was further funded by the Swiss National Science Foundation, the Swiss Nanoscience Institute, the Swiss NCCR QSIT, the ERC Advanced Investigator Grant QUEST and the EU flagship project graphene. M.-H.L. and K.R. acknowledge financial support by the Deutsche Forschungsgemeinschaft (SFB 689).

The authors thank Romain Maurand for fruitful discussions.


\bibliographystyle{unsrt}
\bibliography{mainrefs}

\newpage

\section{Supplementary information}

\begin{figure}[htbp]
    \centering
      \includegraphics[width=0.5\textwidth]{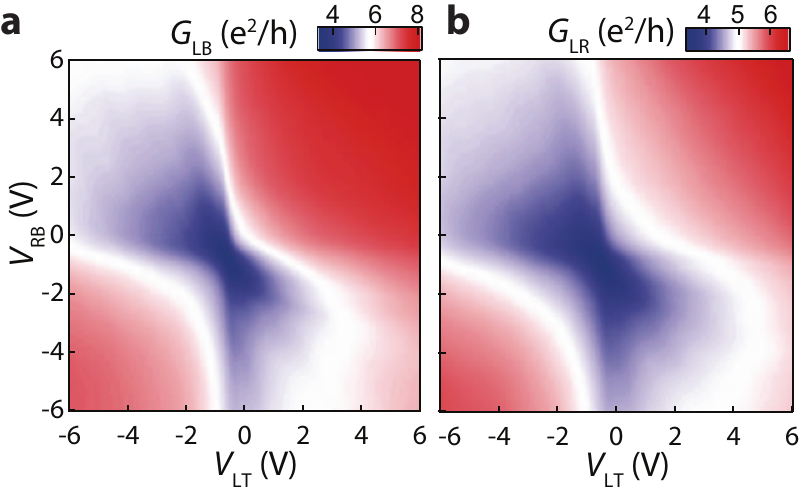}
    \caption{\textbf{a,} $G_{\rm LB}(V_{\rm LT},V_{\rm RB})$, when the current is injected from the L contact. \textbf{b,} $G_{\rm LR}(V_{\rm LT},V_{\rm RB})$. }
    \label{fig:supfig1}
\end{figure}

In SFig.~\ref{fig:supfig1}a) and b) the conductance $G_{\rm LB}(V_{\rm LT},V_{\rm RB})$ and $G_{\rm LR}(V_{\rm LT},V_{\rm RB})$ are given, respectively. The conductance for L to B is higher than L to T in the p-n regime. This can result from more trajectories from L to B than L to R, which have close to perpendicular incidence at the p-n interface. Another reason can be negative refraction, which also bends to trajectories towards the bottom contact.

In SFig.~\ref{fig:supfig1}a) and b) the conductance $G_{\rm LT}(V_{\rm LT},V_{\rm RB})$ and $G_{\rm LB+LR}(V_{\rm LT},V_{\rm RB})$ are given, respectively at $50\,$mT. Similarly, SFig.~\ref{fig:supfig1}c) and d) shows the conductance maps at $100\,$mT.

\begin{figure}[htbp]
    \centering
      \includegraphics[width=0.9\textwidth]{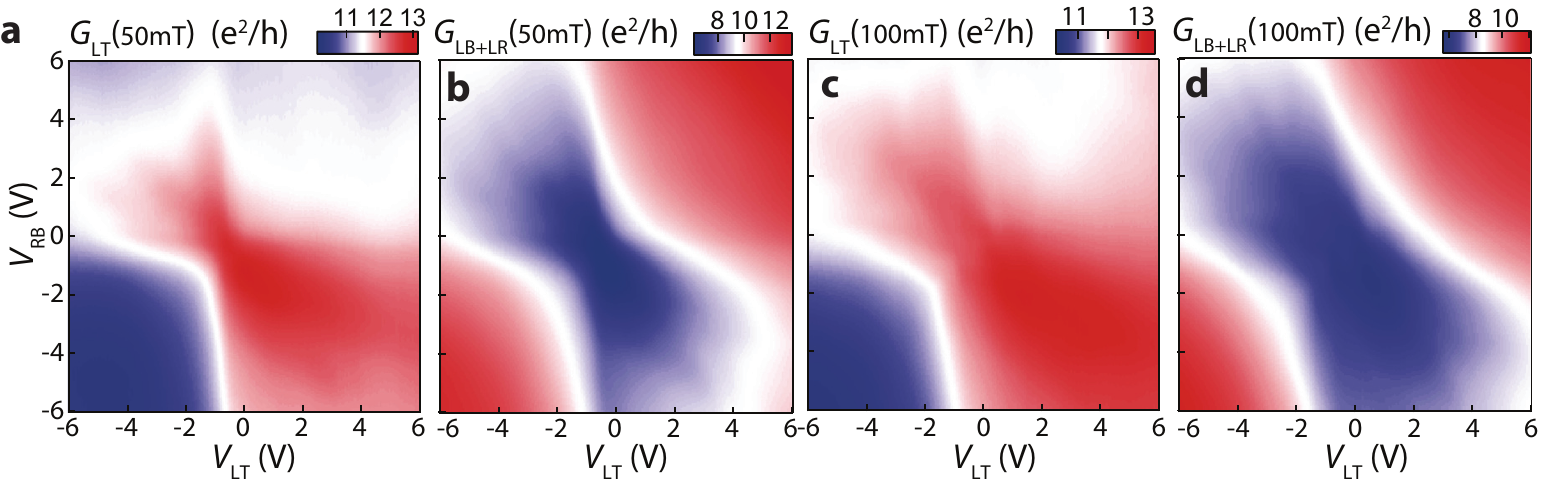}
    \caption{\textbf{a,} $G_{\rm LT}(V_{\rm LT},V_{\rm RB})$ at $50\,$mT. \textbf{b,} $G_{\rm LB+LR}(V_{\rm LT},V_{\rm RB})$ at $50\,$mT. \textbf{c,} $G_{\rm LT}(V_{\rm LT},V_{\rm RB})$ at $100\,$mT. \textbf{d,} $G_{\rm LB+LR}(V_{\rm LT},V_{\rm RB})$ at $100\,$mT.}
    \label{fig:supfig2}
\end{figure}

\end{document}